\documentclass[mathpazo]{cicp}

\begin{document}
%%%%% title : short title may not be used but TITLE is required.
% \title{TITLE}
% \title[short title]{TITLE}
\title{A lattice Boltzmann study of phase separation in\\
liquid-vapor systems with gravity}

%%%%% author(s) :
% single author:
% \author[name in running head]{AUTHOR\corrauth}
% [name in running head] is NOT OPTIONAL, it is a MUST.
% Use \corrauth to indicate the corresponding author.
% Use \email to provide email address of author.
% \footnote and \thanks are not used in the heading section.
% Another acknowlegments/support of grants, state in Acknowledgments section
% \section*{Acknowledgments}
%\author[O.~Author]{Only Author\corrauth}
%\address{School of Mathematical Sciences, Beijing Normal University,
%Beijing 100875, P.R. China}
%\email{{\tt author@email} (O.~Author)}

% multiple authors:
% Note the use of \affil and \affilnum to link names and addresses.
% The author for correspondence is marked by \corrauth.
% use \emails to provide email addresses of authors
% e.g. below example has 3 authors, first author is also the corresponding
%      author, author 1 and 3 having the same address.
% \author[Zhang Z R et.~al.]{Zhengru Zhang\affil{1}\comma\corrauth,
%       Author Chan\affil{2}, and Author Zhao\affil{1}}
% \address{\affilnum{1}\ School of Mathematical Sciences,
%          Beijing Normal University,
%          Beijing 100875, P.R. China. \\
%           \affilnum{2}\ Department of Mathematics,
%           Hong Kong Baptist University, Hong Kong SAR}
% \emails{{\tt zhang@email} (Z.~Zhang), {\tt chan@email} (A.~Chan),
%          {\tt zhao@email} (A.~Zhao)}
% \footnote and \thanks are not used in the heading section.
% Another acknowlegments/support of grants, state in Acknowledgments section
% \section*{Acknowledgments}

\author[A.~Cristea et.~al.]{Artur Cristea\affil{1},
	Giuseppe Gonnella\affil{2},
	Antonio Lamura\affil{3} and\\
	Victor Sofonea\affil{1}\comma\corrauth}
\address{\affilnum{1}\ Center for Fundamental and Advanced Technical Research,
	Romanian Academy,
	Bd. Mihai Viteazul 24, 300223 Timi\c soara, Romania \\
	\affilnum{2}\ Dipartimento di Fisica, Universit\`{a} di Bari
	{\it and} INFN, Sezione di Bari,
	Via Amendola 173, 70126 Bari, Italy \\
	\affilnum{3}\ Istituto Applicazioni Calcolo, CNR,
	Via Amendola 122/D, 70126 Bari, Italy}
\emails{{\tt f1astra@acad-tim.tm.edu.ro} (A.~Cristea),
	{\tt gonnella@ba.infn.it} (G.~Gonnella),
	{\tt a.lamura@ba.iac.cnr.it} (A.~Lamura),
	{\tt sofonea@acad-tim.tm.edu.ro} (V.~Sofonea)}

%%%%% Begin Abstract %%%%%%%%%%%
\begin{abstract}
Phase separation of a two-dimensional  van der Waals fluid subject
to  a gravitational force is studied by numerical simulations based
on lattice Boltzmann methods (LBM) implemented with a   finite
difference scheme. A growth exponent $\alpha=1$ is measured in the
direction of the external force.
\end{abstract}
%%%%% end %%%%%%%%%%%

%%%%% AMS/PACs/Keywords %%%%%%%%%%%
\pac{47.11.-j, 47.20.Hw, 05.70.Ln}
%\ams{52B10, 65D18, 68U05, 68U07}

%%%% maketitle %%%%%
\maketitle

%%%% Start %%%%%%

\section{Introduction}

Phase ordering in fluids is an important   process
that still needs to be completely understood in  many cases of
practical relevance. When a fluid is quenched from an initial
disordered state into a regime of two-phase coexistence below the
spinodal line, domains of the two phases are formed and grow with
time. The typical size $R$ of domains follows the power law $R \sim
t^{\alpha}$ with the growth exponent $\alpha$ being {\it
universal} in the sense that  it does not depend on the microscopic
details of the fluid, assuming only a few values related to the physical
mechanism operating during phase separation \cite{bray}.
Hydrodynamics is in general relevant and  the coupling
with the velocity field can change the value of the growth exponent
$\alpha$ from that of purely diffusive growth \cite{yeomans,onuki}.

In this paper we consider the ordering  of a liquid-vapor  system
subject to  an external field mimicking the effects of gravity. The
role   of gravity on phase ordering has been more studied in binary
systems.  In critical quenches, after an initial diffusive growth
with exponent $1/3$, there is a viscous growth characterized by
$\alpha=1$ followed by an inertial regime with $\alpha=2/3$
\cite{cates}. Gravity becomes relevant when heavy domains resting on
top of light ones become gravitationally unstable, thus accelerating the
domain growth \cite{goldburg}. This occurs at late stages making
inertial growth difficult to observe. A theoretical analysis
neglecting hydrodynamical contributions suggests an exponent
$\alpha_z=1$ for the size of domains in the vertical direction
\cite{binder}. There are   few studies of phase separation for
liquid-vapor systems. In two-dimensional simulations
 the values $\alpha=1/2$ for high viscosity fluids and $\alpha=2/3$
for low-viscosity fluids have been found \cite{yeomans, pre2004}. We are not
aware of simulations made on a liquid-vapor system subject to gravity,
where the growth exponent is measured.

We address this problem by applying the lattice Boltzmann method (LBM)
to simulate a van der Waals fluid described by the Navier-Stokes and
the continuity equation. LBM have been proved successful in studying fluids
with mesoscopic structures (liquid-vapor interfaces in our case) on
large time scales, as it is needed for phase separation
\cite{yeomans,cates,succi1,succi2,succi3,succi4,succi5}. In our
approach, the thermodynamic description is based on a free-energy
functional where interfaces are described at a coarse-grained level.
The free-energy interface cost is expressed, as usual in van der
Waals-Landau models, in terms of gradients of the density field.
Locally, the fluid satisfies the van der Waals state equation.
A finite difference version of LBM is implemented
where
the relationship
$c=\delta s / \delta t$
among the lattice speed $c$ and the space and time
steps $\delta s$ and $\delta t$ does no longer hold, as in standard
{\emph{collision - streaming}} LBM \cite{succi1,succi2,succi3,succi4,succi5}.
The rejection of this condition has two advantages. First, this
allows one to further consider multicomponent fluid systems
where the masses of the component particles, as well as the lattice speeds,
may be no longer identical \cite{cejp,galerkin}. Second, higher order
numerical schemes (including flux limiter schemes) may be considered
in order to reduce unphysical effects like the spurious velocity and the
numerical viscosity
\cite{pre2004,cejp,galerkin,ijmpc2003,jcph2003,ijmpc2005,mcsim2006}.
The use of high order numerical schemes in finite difference LBM
helps further to improve the numerical stability and accuracy
\cite{pre2004,ijmpc2005} while providing a convenient alternative
to interpolation supplemented LBM \cite{he1,he2}.

Our main results is that the  sedimentation process induced by
gravity is characterized by an exponent $\alpha=1$ independently on
the values of viscosity and gravity.

The paper is organized as follows. Our LBM approach is described in
Section 2; numerical results are shown in Section 3 and conclusions
will be drawn in Section 4.

\section{Description of the model}

In this paper, we use the D2Q9 isothermal finite difference lattice Boltzmann 
model
in two dimensions,  which is well known in the literature
\cite{succi2,succi3,succi4,ijmpc2003,qianepl}. This model relies on the 
following set of
${\mathcal{N}}=9$ evolution equations for the non-dimensionalized
distribution functions $f_{i}({\mathbf{r}},t)$, $i = 0,\,1,\,\ldots,\,
{\mathcal{N}}-1$, defined
in the nodes ${\mathbf{r}}=(x,y)$ of a lattice with
$\Lambda_x \times \Lambda_y$ nodes
\cite{succi1,succi2,succi3,succi4,succi5,qianepl,cao,pre2004,mcsim2006}
\begin{eqnarray}
f_{i}({\mathbf{r}},t+\delta t) & = & f_{i}({\mathbf{r}},t)\,-\,
\delta t\,{\mathbf{e}}_{i}\cdot\nabla\,f_{i}({\mathbf{r}},t)\,-\,
\frac{\,\delta t\,}{\,\tau\,}\,\left[\,f_{i}({\mathbf{r}},t)\,-\,f^{eq}_{i}
({\mathbf{r}},t)\,\right]
\nonumber\\
& + & \frac{\,\delta t\,}{\,\chi c^{2}\,}\,{\mathbf{F}}\cdot\,\left[\,
{\mathbf{e}}_{i}\,-\,
{\mathbf{u}}({\mathbf{r}},t)\,\right]\,f^{eq}_{i}({\mathbf{r}},t) . \rule{0mm}
{7mm}
\label{lbevol}
\end{eqnarray}
In this non-dimensionalized model, the mass of particles equals 1.
To reduce numerical errors,
the second order Monitorized-Central-Difference flux limiter scheme
\cite{pre2004,mcsim2006,cejp,leveque,pre2007}
was used to calculate the space derivative $\nabla f_{i}({\mathbf{r}},t)$.

The local values of the fluid quantities (particle number density $n$ and
velocity ${\mathbf{u}}$) are derived from the distribution functions, as
follows
\begin{eqnarray}
n({\mathbf{r}},t) & = & \sum_{i=0}^{\mathcal{N}-1}\,f_{i}({\mathbf{r},t}) , \\
{\mathbf{u}}({\mathbf{r}},t) & = & \frac{\,1\,}{\,n({\mathbf{r}},t)\,}\,
\sum_{i=0}^{\mathcal{N}-1}\,f_{i}({\mathbf{r},t})\,{\mathbf{e}}_{i} .
\rule{0mm}{7mm}
\end{eqnarray}
The velocity vectors $\{\,{\mathbf{e}}_{i}\,\}$ are given by
\begin{eqnarray}
{\mathbf{e}}_{0} & = & 0 \nonumber\\
{\mathbf{e}}_{i} & = & \left(\,\cos\frac{\,\pi(i-1)\,}{\,2\,}\ ,\ 
\sin\frac{\,\pi(i-1)\,}{\,2\,}\,\right)\,c
\quad\qquad \,\quad\,\,  {\mathrm{for}}\quad i\,=\,1,\,\ldots\,,\,4 \rule{0mm}
{7mm}\\
{\mathbf{e}}_{i} & = & \left(\,\cos\frac{\,\pi(2i-9)\,}{\,4\,}\ ,\ 
\sin\frac{\,\pi(2i-9)\,}{\,4\,}\,\right)\,
c\sqrt{2}\quad \,\quad {\mathrm{for}}\quad i\,=\,5,\,\ldots\,,\,8 \rule{0mm}
{7mm}
\nonumber
\end{eqnarray}
where $c\,=\,c/c_{R}=\sqrt{\theta/\chi}$ is a non-dimensionalized speed,
$\theta=T/T_{R}$ is the non-dimensio\-na\-lized system 
temperature, and $\chi\,=\,1/3$.
As discussed in \cite{ijmpc2003,artur},
the following reference quantities $n_{R}=N_{A}/V_{mc}$,
$T_{R}=T_{c}$ and $c_{R}=\sqrt{k_{B}T_{c}/m_R}$, where $N_{A}$ is Avogadro's
number, $V_{mc}$ is the molar volume at the critical point of temperature 
$T_{c}$,
$m_{R}$ is the mass of the fluid particles
and $k_{B}$ is Boltzmann's constant, may be used to get the 
non-dimensionalized 
values of the particle number density, temperature and speed, respectively. 
The system size is chosen as reference length $l_R$, the reference
quantities $t_{R}$ and $a_{R}$ for time and acceleration follow from
\begin{equation}
\frac{\,t_{R}c_{R}\,}{\,l_{R}\,}\,=\,1 \qquad ,
\qquad \frac{\,a_{R}t_{R}\,}{\,c_{R}\,}\,=\,1 .
\end{equation}
Since we use finite difference schemes 
\cite{ijmpc2003,cejp,pre2004,cao,leveque}
to evolve the particle distribution functions
according to Eq.(\ref{lbevol}), the lattice spacing $\delta s$ and the time 
step $\delta s$
are no longer related to the lattice speed $c$ as in standard lattice 
Boltzmann models
\cite{succi1,succi2,succi3,succi4}, the temperature $\theta$ is now a control 
parameter
in our simulations. This feature of the  finite difference approach allows us 
to change 
the system  temperature $\theta$ (and hence also the lattice speed $c$) while 
preserving
the lattice spacing $\delta s$ and the system size in each direction (i.e., 
the 
corresponding number of lattice nodes). In such models there is more freedom
to choose the discrete velocity set, as done recently in a thermal model
\cite{watari} where the possibility of having different sets of velocities
allows to release the constraint of constant temperature. 

The equilibrium distribution functions $f^{eq}_{i}=f^{eq}_{i}({\mathbf{r}},t)$
that appear in the evolution equation (\ref{lbevol}) of the D2Q9 model
are expressed as series expansion up to second order with
respect to the fluid velocity ${\mathbf{u}}$ 
\cite{succi2,succi3,succi4,qianepl}
\begin{equation}
f^{eq}_{i}\,=\,w_{i}n\,\left[\,1\,+\,\frac{\,{\mathbf{e}}_{i}
\cdot{\mathbf{u}}\,}{\,\chi c^{2}\,}
\,+\,\frac{\,({\mathbf{e}}_{i}\cdot{\mathbf{u}})^{2}\,}{\,2\chi^{2}c^{4}\,}\,
-\,
\frac{\,({\mathbf{u}})^{2}\,}{\,2\chi c^{2}\,}\,\right]
\end{equation}
where the weight coefficients are
\begin{equation}
w_{i}\,=\,\left\{\,\begin{array}{ccl}
\frac{\,4\,}{\,9\,} & , & i\,=\,0\\
\frac{\,1\,}{\,9\,} & , & i\,=\,1,\,\ldots\,4 \rule{0mm}{6mm}\\
\frac{\,1\,}{\,36\,} & , & i\,=\,5,\,\ldots\,8 \rule{0mm}{6mm}\\
\end{array}\right.
\end{equation}

The force ${\mathbf{F}}$ in Eq.~(\ref{lbevol}) is introduced to recover the 
macroscopic 
equations of an isothermal van der Waals fluid subjected to the gravitational 
acceleration
${\mathbf{a}}$. When using the reference pressure $p_{R}=mn_{R}c^{2}_{R}$,
this force has the Cartesian components 
\cite{pre2004,ijmpc2003,artur,he3,he4,he5}
\begin{equation}
F_{\alpha}\,=\,\frac{\,1\,}{\,n\,}\,\partial_{\alpha}\,(\,p^{ideal}\,-\,
p^{waals}\,)\,+\,\kappa\,\partial_{\alpha}\,(\nabla^{2}n)\,+\,a_{\alpha}
\label{defor}
\end{equation}
where
\begin{equation}
p^{ideal}\,=\,n\theta
\end{equation}
is the ideal gas pressure and
\begin{equation}
p^{waals}\,=\,\frac{\,n\theta\,}{\,1\,-\,b n\,}\,-\, a \,n^{2}
\end{equation}
is the van der Waals fluid pressure in non-dimensionalized form,
$\kappa$ is a constant which controls the
surface tension and ${\mathbf{a}}$ is the gravitational acceleration.
The parameters $a$ and $b$ are given by
\begin{eqnarray}
a&=&\frac{9}{8} \frac{\theta_c}{n_c} \\ 
b&=&\frac{1}{3 n_c}
\end{eqnarray}
where $n_c$ is the particle density at the critical temperature $\theta_c=1$.
In the following we will consider $n_c=1$ so the the
van der Waals fluid pressure reads
\begin{equation}
p^{waals}\,=\,\frac{\,3 n\theta\,}{\,3\,-\, n\,}\,-\, \frac{9}{8} \,n^{2} .
\end{equation}
Note that the force (\ref{defor}) was already used in
\cite{pre2004,ijmpc2003,mcsim2006,artur} to 
investigate the dynamics and morphology of phase separation in liquid-vapour
systems in the absence of gravity (${\mathbf{a}}=0$). 

A Chapman-Enskog expansion shows that the continuity
and Navier-Stokes equations are recovered in the continuum limit:
\begin{equation}
\partial_t n + \partial_{\beta} (n u_{\beta})=0 ,
\end{equation}
\begin{equation}
\partial_t (n u_{\alpha})  + \partial_{\beta} (n u_{\alpha} u_{\beta})
= -\partial_{\alpha} p^{waals}
+ \kappa n \partial_{\alpha}(\nabla^2 n)
+ n a_{\alpha}
+ \nu \partial_{\beta}
[n(\partial_{\alpha} u_{\beta} + \partial_{\beta} u_{\alpha})]
\end{equation}
with kinematic viscosity $\nu= \theta \tau$.
The terms $-\partial_{\alpha} p^{waals}
+ \kappa n \partial_{\alpha}(\nabla^2 n)$ at the r.h.s. of the
Navier-Stokes equation can also be written
in the form $-\partial_{\beta} P_{\alpha \beta}$ where
the pressure tensor $P_{\alpha \beta}$ is related to the free energy
functional of the van der Waals fluid
\cite{rowl}
\begin{equation}
\Psi = \int d {\bf r} \left [
\psi(n,\theta) + \frac{\kappa}{2} (\nabla n)^2
\right ] ,
\end{equation}
$\psi(n,\theta)$ being the bulk free energy density
\begin{equation}
\psi(n,\theta)=n \theta \ln (\frac{3 n}{3 -n})-\frac{9}{8} n^2 .
\end{equation}
The pressure tensor is \cite{evans}
\begin{equation}
P_{\alpha \beta} = p \delta_{\alpha \beta} + \kappa \partial_{\alpha} n
\partial_{\beta} n
\end{equation}
with
\begin{equation}
p = p^{waals} - \kappa n \nabla^2 n -\frac{\kappa}{2} (\nabla n)^2
\end{equation}
where $p^{waals}=n \psi^{'}(n)-\psi$ is the equation of state
with the critical point at $n_c=1$, $\theta_c=1$.

In the sequel, we will consider the case of an acceleration directed
upwards: $a_{x}\,=\,0$, $a_{y}\,=\,g$. Periodical boundary
conditions
were considered in the horizontal direction and standard bounce back boundary
conditions \cite{succi1,succi2,succi3,succi4,succi5}
were imposed on top and bottom walls.

\section{Numerical results}

In this Section we report the results of our simulations.
For runs we used either a square lattice with $\Lambda_x=\Lambda_y=1024$ or
a rectangular one with $\Lambda_x=512$ and 
$\Lambda_y=4096$, lattice spacing
$\delta s = 1/256$ and time step $\delta t = 10^{-5}$, as in \cite{pre2004}.
All quenches below the critical temperature $\theta_c=1$ were to the
temperature $\theta = 0.79$
where the coexisting densities are $n_{liquid} = 1.956$ and
$n_{vapor} = 0.226$.
Each simulation was started with small fluctuations ($0.1 \%$)
in the density about
a mean value ${\hat n}$ that was either symmetric (${\hat n} = 1.09$,
liquid fraction $\beta=0.5$) or slightly off-symmetric (${\hat n} = 1.0$,
$\beta=0.45$). The parameter $\kappa$ controlling the
surface tension was set to $5 \times 10^{-6}$ to have an interface
thickness of $\sim 6 $ lattice spacings. 
The corresponding value of the surface tension $\sigma$ was evaluated by using
the Laplace law \cite{artur} 
and a value $\sigma = 2.0 \times 10^{-3}$ was measured.
The value of the
constant $g$ controlling the
external acceleration,
was varied in the range $[0.0001,0.005]$. Results were not
dependent on its specific value as shown in the following.    
The viscosity was varied by changing $\tau$.
We used the values $\tau = 10^{-4}$ and $\tau = 10^{-3}$,
which allow us to access low and high viscosity regimes,
respectively, as shown in a previous work where the present model
without gravity was used to study the phase separation in liquid-gas
systems \cite{pre2004}.

The process of phase separation depends on the interplay among three
main driving forces:
The viscous one $F_v = n \nu l_R c_R$, the gravity one $F_g = n g l_R^3$, 
and the
surface tension one $F_s = \sigma l_R$. It may be then useful to
evaluate their relative contributions
introducing the Bond number
$\displaystyle Bo = \frac{F_g}{F_s} = \frac{n g l_R^2}{\sigma}$, 
the Capillary number
$\displaystyle Ca = \frac{F_v}{F_s} = \frac{n \nu c_R}{\sigma}$, and
the ratio $\displaystyle \frac{Bo}{Ca} = \frac{F_g}{F_v} = \frac{g l_R^2}
{\nu c_R}$.
The choice of the input parameters is such to access the ranges 
$0.05 \leq Bo \leq 2.5$,
$0.04 \leq Ca \leq 0.4$, and $ 0.13 \leq Bo/Ca \leq 65$.

\begin{figure}
\includegraphics[width=0.8\textwidth]{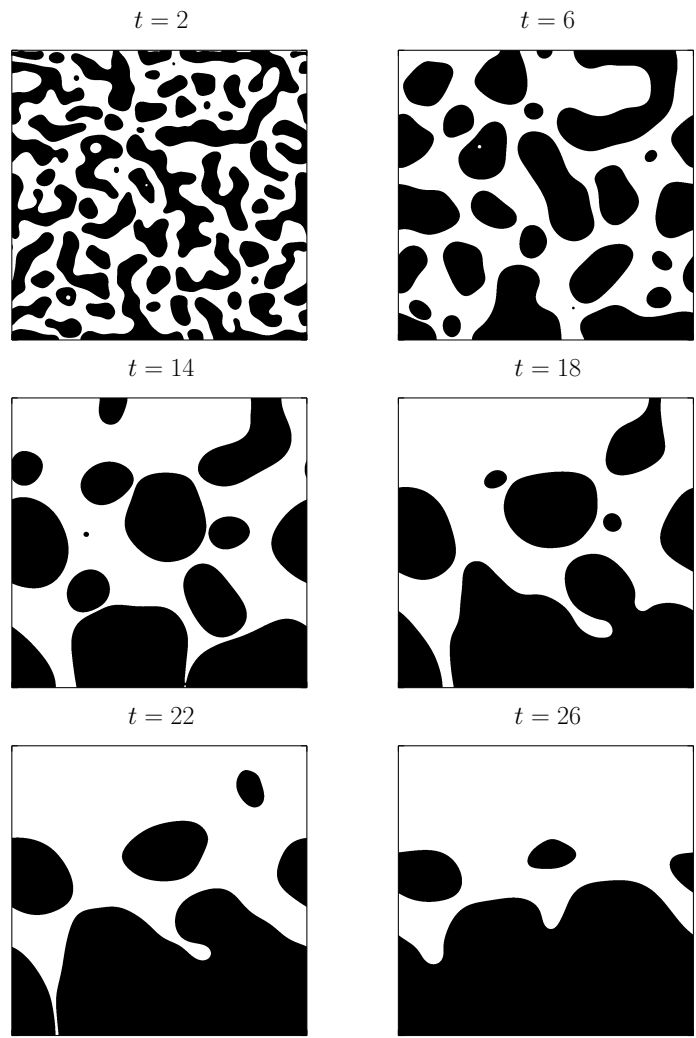}
\caption{Contour plots of the density $n$ in the case with
$\tau=10^{-3}$, $g=0.005$ and $\beta=0.5$.
Color code: white/black $\rightarrow$ liquid/vapor.
\label{fig1}}
\end{figure}

\begin{figure}
\includegraphics[width=0.8\textwidth]{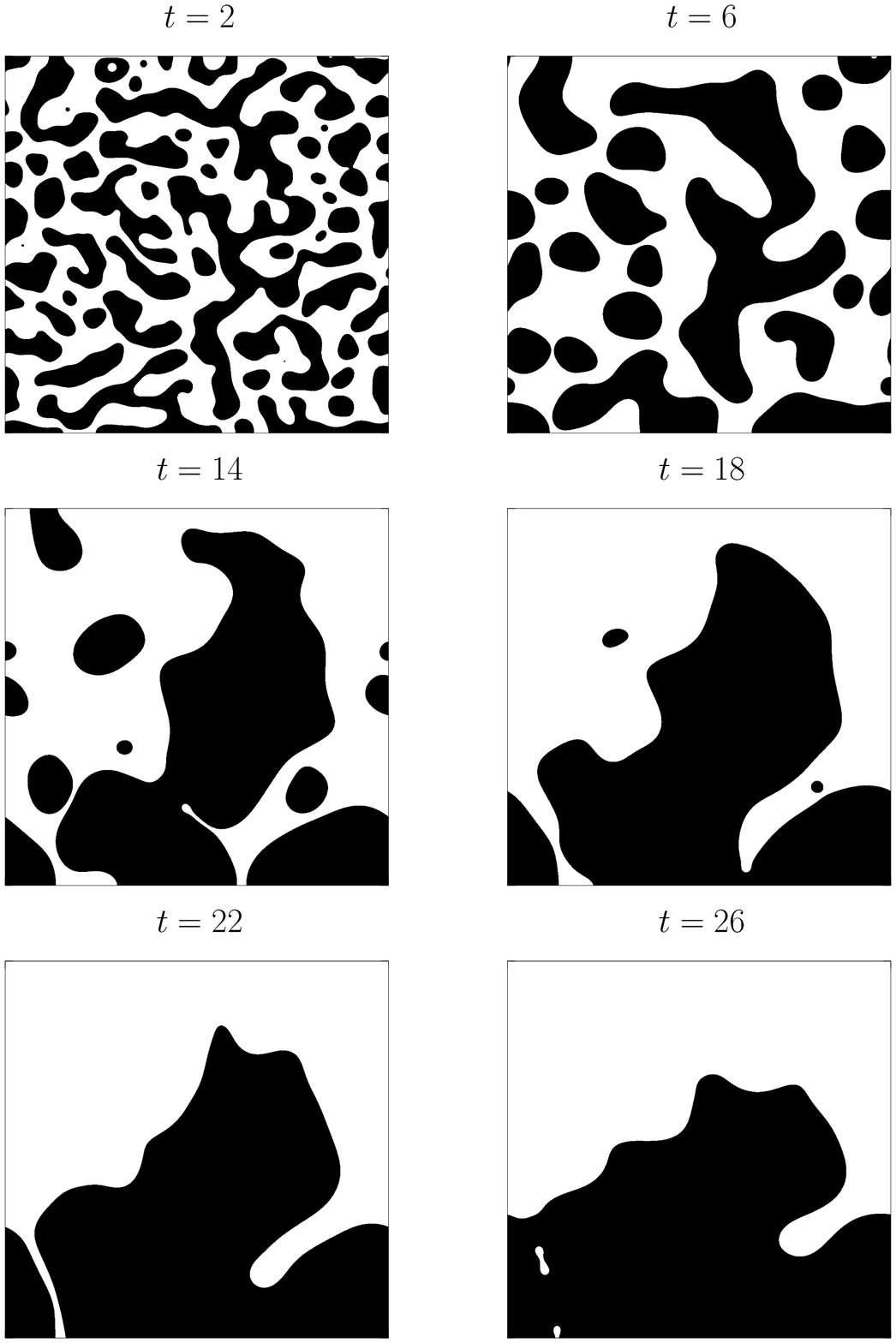}
\caption{Contour plots of the density $n$ in the case with
$\tau=10^{-4}$, $g=0.005$ and $\beta=0.5$.
Color code: white/black $\rightarrow$ liquid/vapor.
\label{fig2}}
\end{figure}

After the initial stages when the mixture starts to phase separate,
the effect of the gravitational force is to accumulate material at walls:
The heavy phase (liquid) is moved to the top wall
and the light phase (vapor) stays
at the
bottom wall. The evolution of domains in the cases
at high ($\tau=10^{-3}$) and low ($\tau=10^{-4}$) viscosities is shown
in Figs.~\ref{fig1}-\ref{fig2} for $g=0.005$. 
At intermediate times between $t \simeq 3$
and $t \simeq 10$, anisotropic patterns can be observed in the bulk
region far from the walls, with
domains slightly elongated along the vertical direction.
The main difference that can be observed between the two figures
is the presence of many droplets
in the case at high viscosity as compared to the case at low viscosity.
The reason is due to the fact that in the latter case hydrodynamics is
effective in coalescing droplets, thus producing a more homogeneous pattern.

\begin{figure}
\includegraphics[width=0.55\textwidth]{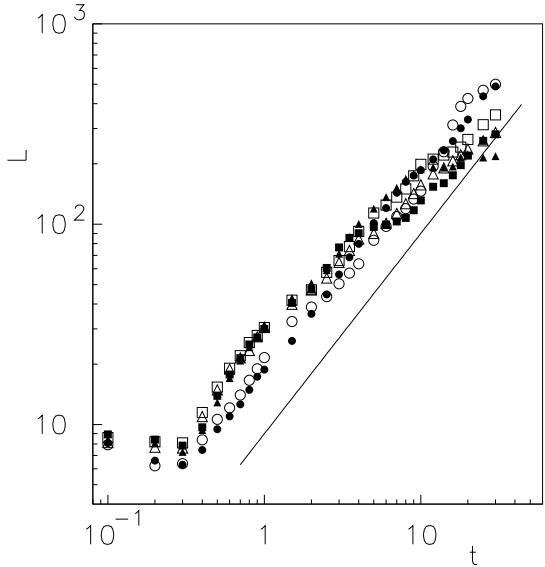}
\caption{Time evolution of the average thickness of layers at walls
in the cases with $g=0.0001 (\triangle), 0.001 (\Box), 0.005 (\circ)$
for $\tau= 10^{-3}$ (filled symbols),\  $10^{-4}$ (empty symbols) 
and $\beta=0.5$.
The solid line is a guide to
the eye and has slope $1$.
\label{fig3}}
\end{figure}

In order to gain some insight into the law governing the accumulation
of material at walls, we measured the average thickness $L$ of layers
adjacent to the walls.
For each column of the lattice we looked for all the sites next to the walls
where there was an interface between the liquid and vapor phases. To be more
specific we found all the lattice sites along the $x$-direction
with the smallest distance
$y^*_b(x)$ to the bottom wall
such that
$[n(x,y^*_b(x))-{\hat n}][n(x,y^*_b(x)+1)-{\hat n}]<0$ with
$y^*_b(x) < \Lambda_y/2$
and all the sites with the smallest distance $\Lambda_y - y^*_t(x)$
to the top wall such that
$[n(x,y^*_t(x))-{\hat n}][n(x,y^*_t(x)-1)-{\hat n}]<0$
with $y^*_t(x) > \Lambda_y/2$.
We defined
\begin{equation}
L = \frac{1}{2 \Lambda_x} \sum_{x=1}^{\Lambda_x}[y^*_b(x)+(\Lambda_y-y^*_t(x))]
< \frac{\Lambda_y}{2} .
\end{equation}
The time evolution of $L$ is reported in Fig.~\ref{fig3} for different values
of $\tau$ and $g$. We have a clear
and convincing indication that in all the cases the growth
is consistent with a power law with growth exponent $\alpha=1$ which 
depends neither on $g$ nor on $\tau$.
The growth is observed over almost two time decades until the system
is entirely separated in two parts of different composition.
In particular, we want to stress the fact that the exponent
$\alpha=1$ is observed in both cases,
when the gravitational force $F_g$ is small
compared to the surface tension force $F_s$ ($Bo < 1$), as well as
when $F_g$ is small compared to the viscous one $F_v$ ($Bo/Ca < 1$).
This indicates that the
existence of the same scaling exponent $\alpha$ in the gravity direction
is exclusively due to the presence of the gravity force $F_g > 0$.
Our result is in agreement with previous studies of mixtures where 
hydrodynamics was
neglected \cite{binder,lacasta,puri1,puri2}.
In another study
on the phase separation of binary fluids \cite{maritan} where hydrodynamic
effects
were considered, it was argued that
the growth exponent is $\alpha=0.6\pm 0.1$ and
is not affected by the presence of gravity.
The present study shows that the value of the growth exponent
is independent on the value of the viscosity and of the gravity.
Similar results were obtained when considering the case of a slightly
off-symmetric mixtures with $\beta=0.45$.

\begin{figure}
\includegraphics[width=0.55\textwidth]{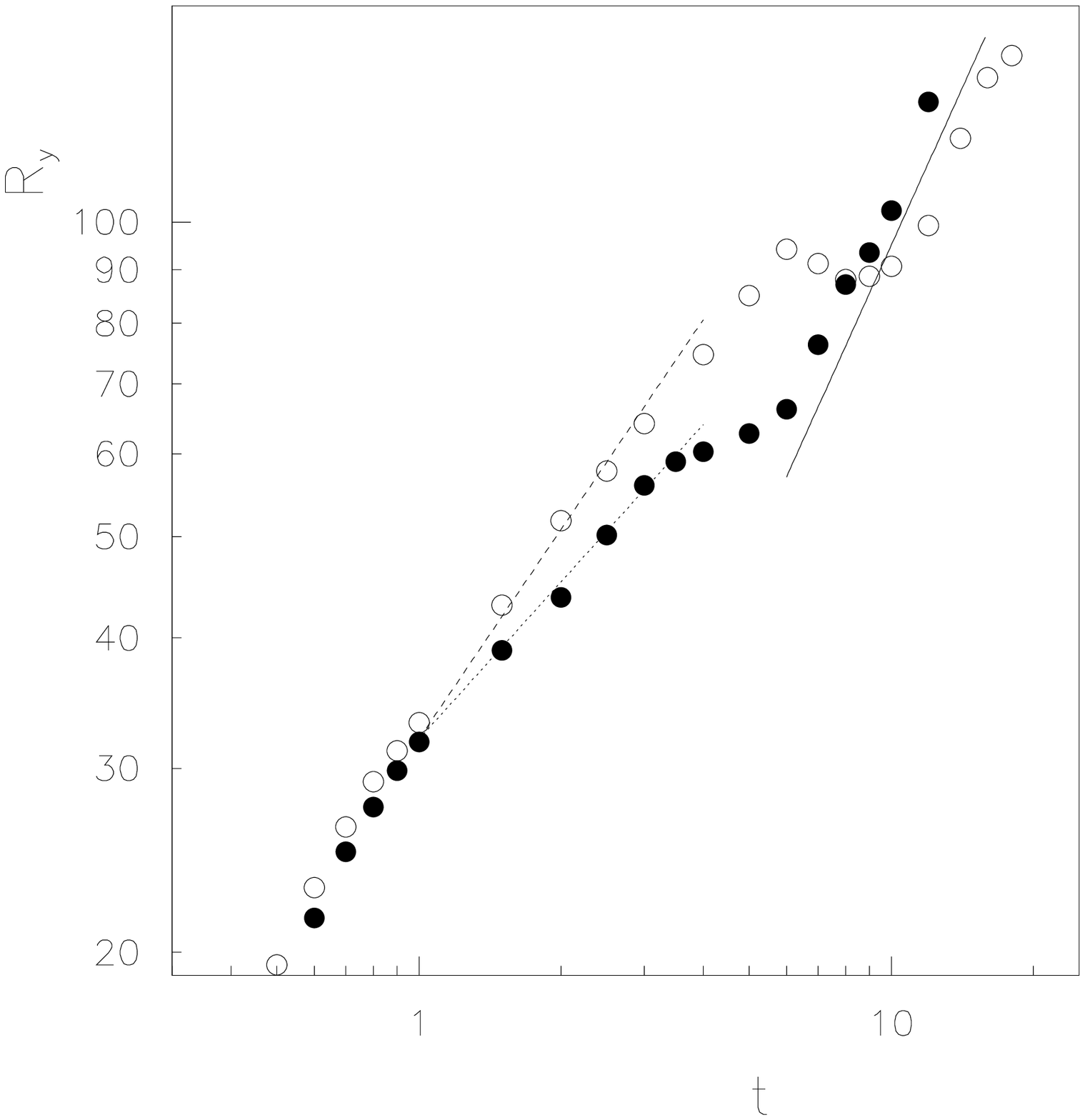}\\*
\includegraphics[width=0.55\textwidth]{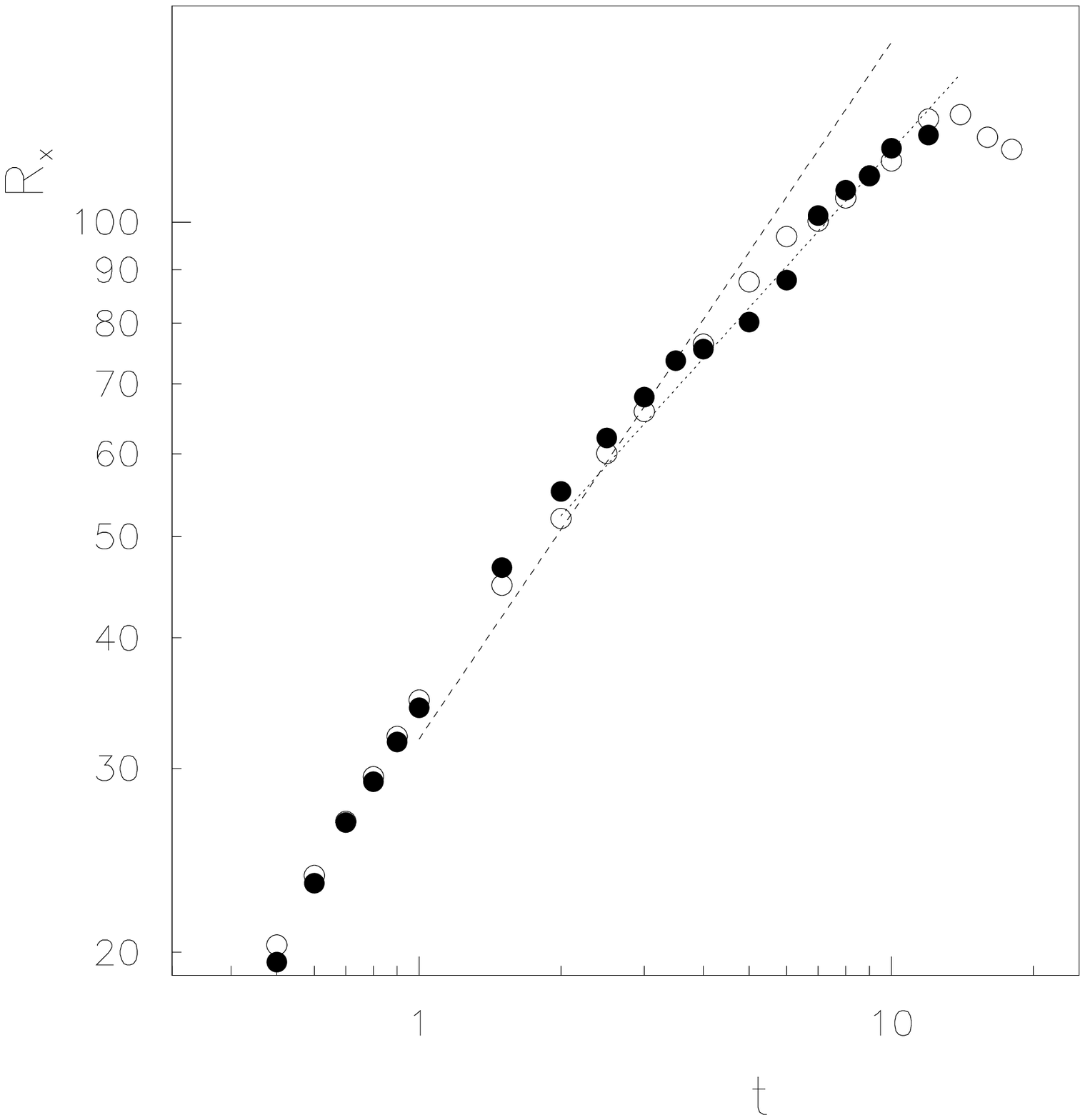}
\caption{Time evolution of the size of domains $R_y$ (upper panel) and 
$R_x$ (lower panel)
in the case with $g=0.005$
for $\tau= 10^{-3}$ ($\bullet$),\  $10^{-4}$ ($\circ$) 
and $\beta=0.5$ on a lattice $512 \times 4096$.
The lines serve as eyeguide and have the slopes $1/2$ (dotted line),
$2/3$ (dashed line), and $1$ (full line).
\label{fig4}}
\end{figure}

In order to better characterize the morphology of domains we simulated
the behaviour of a very large system with symmetric composition ($\beta=0.5$)
and size $512 \times 4096$, for $g=0.005$ and $\tau\in\{10^{-4},\,10^{-3}\}$.
Also in this case we found that $L$ grows with the exponent $1$.
In order to estimate the domains size in the two spatial directions, due
to the anisotropy induced by gravity,
we computed the inverse of the first moment of the structure factor
\cite{lam1,lam2}
\begin{equation}
R_x(t)= \pi \frac{\int d\mathbf{k} C(\mathbf{k},t)}{\int d\mathbf{k} k_x
C(\mathbf{k},t)}
\end{equation}
and similarly for the $y$ direction where
\begin{equation}
C(\mathbf{k},t) = <{\tilde n}(\mathbf{k},t) {\tilde n}(-\mathbf{k},t)>
\end{equation}
with ${\tilde n}(\mathbf{k},t)$ the spatial Fourier transform of 
$n(\mathbf{r},t)-{\hat n}$.
The results are shown in Fig.~\ref{fig4}. Along the gravity direction,
we see that $R_y$ starts to grow with the same exponents as in the case
 without gravity:
$1/2$ for $\tau=0.001$ and $2/3$ for $\tau=0.0001$ \cite{pre2004}.
The transition to a regime consistent 
with the growth exponent $1$ is observed at later times. Along
the horizontal direction, we get $R_x \sim t^{1/2}$ in the
high viscosity case ($\tau=0.001$), a behaviour that is similar to the case
when the system is subjected to no gravity. However, at low viscosity 
($\tau=0.0001$) the growth exponent along the horizontal direction
in the presence of gravity
is smaller than the expected value $2/3$, which is achieved without gravity.
It is interesting to note that similar conclusions were drawn in the case
of phase separation of binary mixtures under gravity \cite{badal}. In their
study, the authors observed that in the diffusive and viscous regimes the
growth exponent is always equal to
$1$ along the vertical (gravity) direction, while in the horizontal direction
there is a slowing down of the growth rate with respect to the case without
gravity \cite{badal}.
In order to better elucidate these features one would need to 
perform higher resolution simulations to access a wider range of length scales.

\section{Conclusions}

In this paper we have introduced an external gravitational force in an
isothermal lattice Boltzmann model for the van der Waals fluid. We have studied
phase separation in systems with different viscosities and various values
of the gravitational acceleration.
In the absence of gravity, the growth exponent is known to have specific
values \cite{pre2004}, which depend on the fluid viscosity. When the
liquid-vapor system was subjected to the gravitational force, we
measured the evolution of the characteristic size
(along the gravity direction) of the growing domains and found
the same exponent $\alpha=1$ for all the cases considered, even 
if the fluid viscosity and the gravitational acceleration were different.
Further extension of our parallel computing code to three dimensions
would allow to evaluate the growth exponents in a more realistic case.

%%%% Acknowledgments %%%%%%%%

\section*{Acknowledgments}
The authors acknowledge support from CEEX 11 (2005-2008), as well as
CNR-INFM  for a grant at CINECA Consortium for Supercomputing.

%%%% Bibliography  %%%%%%%%%%


\begin{thebibliography}{99}

\bibitem{bray}
A. J. Bray, Theory of phase-ordering kinetics,
Advances in Physics {\bf 43} (1994) 357 - 459.

\bibitem{yeomans}
J. M. Yeomans, Phase ordering in Fluids, in Annual Reviews of Computational
Physics  {\bf VII}, pp.~61 - 84
(D.Stauffer, Editor), World Scientific, Singapore, (2000).

\bibitem{onuki}
A. Onuki, Phase transitions of fluids in shear flow,
Journal of Physics -- Condensed Matter  {\bf 9} (1997) 6119 - 6157.

\bibitem{cates}
V. M. Kendon, M. E. Cates, I. Pagonabarraga, J.-C. Desplat, and P. Bladon,
Inertial effects in three-dimensional spinodal decomposition of a
symmetric binary fluid mixture: a lattice Boltzmann study,
Journal of Fluid Mechanics {\bf 440} (2001) 147 - 203.

\bibitem{goldburg}
C. K. Chan and W. I. Goldburg,
Late-stage phase separation and hydrodynamic flow in a binary liquid mixture,
Physical Review Letters {\bf 58} (1987) 674 -677.

\bibitem{binder}
S. Puri, K. Binder, and S. Dattagupta,
Dynamical scaling in anisotropic phase-separating systems in a
gravitational field,
Physical Review B {\bf 46} (1992) 98 - 105.

\bibitem{pre2004}
V. Sofonea, A. Lamura, G. Gonnella, and A. Cristea,
Finite-difference lattice Boltzmann model with flux limiters
for liquid-vapor systems, Physical Review E  {\bf 70} (2004) 046702.

\bibitem{succi1}
R. Benzi, S. Succi, and M. Vergassola, The lattice Boltzmann equation:
theory and applications, Physics Reports {\bf 222} (1992) 145 - 197.

\bibitem{succi2}
S. Chen and G. D. Doolen, Lattice Boltzmann Method for fluid Flows,
Annual Review of Fluid Mechanics {\bf 30} (1998) 329 - 364.

\bibitem{succi3}
D. A. Wolf--Gladrow,
Lattice Gas Cellular Automata and Lattice Boltzmann Models,
Springer, Berlin, 2000.

\bibitem{succi4}
S. Succi, The Lattice Boltzmann Equation for Fluid Dynamics and Beyond,
Clarendon Press, Oxford, 2001.

\bibitem{succi5}
B. Duenweg and A. J. C. Ladd,
Lattice Boltzmann simulations of soft matter systems,
in Advences in Polymer Science {\bf 221} (2009) 89 - 166.

\bibitem{cejp}
A. Cristea and V. Sofonea, Two component lattice Boltzmann model
with flux limiters, Central European Journal of Physics
{\bf 2} (2004) 382 - 396.

\bibitem{galerkin}
V.Sofonea, Discontinuous Galerkin schemes for isothermal lattice
Boltzmann models in one dimension, International Journal of Modern
Physics C {\bf 19} (2008) 677 - 688.

\bibitem{ijmpc2003}
A. Cristea and V. Sofonea, Reduction of spurious velocity in finite
difference lattice Boltzmann models for liquid-vapor systems,
International Journal of Modern Physics C {\bf 14} (2003) 1251 - 1266.

\bibitem{jcph2003}
V. Sofonea and R. F. Sekerka, Viscosity of finite difference lattice
Boltzmann models, Journal of Computational Physics {\bf 184} (2003)
422 - 434.

\bibitem{ijmpc2005}
V. Sofonea and R. F. Sekerka, Diffusivity of two-component isothermal
finite difference lattice Boltzmann models, International Journal
of Modern Physics C {\bf 16} (2005) 1075 - 1090.

\bibitem{mcsim2006}
A. Cristea, G. Gonnella, A. Lamura, and V. Sofonea,
Finite-difference lattice Boltzmann model for liquid-vapor systems,
Mathematics and Computers in Simulation {\bf 72} (2006) 113 - 116.

\bibitem{he1}
X. He, L. Luo, and M. Dembo,
Some progress in lattice Boltzmann method. part I. Nonuniform
mesh grids, Journal of Computational Physics {\bf 129} (1996) 357 - 363.

\bibitem{he2}
X. He, Error analysis for the interpolation-supplemented lattice-Boltzmann
equation scheme, International Journal of Modern Physics C
{\bf 8} (1997) 737 - 745.

\bibitem{qianepl}
Y. H. Qian, D. d'Humi\`{e}res, and P. Lallemand,
Lattice BGK Models for Navier-Stokes Equation,
Europhysics Letters {\bf 17} (1992) 479 - 484.

\bibitem{cao}
N. Cao, S. Chen, S. Jin, and D. Mart\'\i nez,
Physical symmetry and lattice symmetry in the lattice Boltzmann method,
Physical Review E {\bf 55} (1997) R21 - R24.

\bibitem{leveque}
R. J. LeVeque, Numerical Methods for Conservation Laws,
Birkh\"auser, Basel, 1992.

\bibitem{pre2007}
G. Gonnella, A. Lamura, and V. Sofonea, Lattice Boltzmann simulation of
thermal nonideal fluids, Physical Review E  {\bf 76} (2007) 036703.

\bibitem{artur}
A. Cristea, Numerical effects in a finite difference lattice Boltzmann model 
for
liquid-vapour systems, International Journal of Modern Physics C
{\bf 17} (2006) 1191 - 1201.

\bibitem{watari}
M. Watari and M. Tsutahara, Two-dimensional thermal model of the 
finite-difference lattice Boltzmann method with high spatial isotropy, 
Physical Review E  {\bf 67} (2003) 036306.

\bibitem{he3}
X.Y.He, X.W.Shan and G.D.Doolen, Discrete Boltzmann equation for nonideal
gases, Physical Review E {\bf 57} (1998) R13 - R16.

\bibitem{he4}
X. Y. He, S. Y. Chen and R. Y. Zhang, A lattice Boltzmann scheme for 
incompressible
multiphase flow and its applications in simulation of Rayleigh - Taylor 
instability,
Journal of Computational Physics {\bf 152} (1999) 642 - 663.

\bibitem{he5}
X. Y. He and G. D. Doolen, Thermodynamic foundations of kinetic theory and
lattice Boltzmann models for multiphase flows, Journal of Statistical Physics 
{\bf 107} 
(2002) 309 - 328.

\bibitem{rowl}
J. S. Rowlinson and B. Widom, Molecular Theory of Capillarity,
Clarendon Press, Oxford, 1982.

\bibitem{evans}
R. Evans, 	
The nature of the liquid-vapour interface and other topics in the
statistical mechanics of non-uniform, classical fluids,
Advances in Physics {\bf 28} (1979) 143 - 200.

\bibitem{lacasta}
A. M. Lacasta, A. Hernandez-Machado, and J. M. Sancho,
Front and domain growth in the presence of gravity,
Physical Review B {\bf 48} (1993) 9418 - 9425.

\bibitem{puri1}
S. Puri, N. Parekh, and S. Dattagupta,
Phase ordering dynamics in a gravitational field,
Journal of Statistical Physics {\bf 75} (1994) 839 - 857.

\bibitem{puri2}
S. Puri, Phase separation kinetics in anisotropic systems,
Physica A {\bf 224} (1996) 101 - 112.

\bibitem{maritan}
W. Ma, A. Maritan, J. R. Banavar, and J. Koplik,
Dynamics of phase separation of binary fluids,
Physical Review A {\bf 45} (1992) R5347 - R5350.

\bibitem{lam1}
F. Corberi, G. Gonnella, and A. Lamura,
Spinodal decomposition of binary mixtures in uniform shear flow,
Physical Review Letters {\bf 81} (1998) 3852 - 3855.

\bibitem{lam2}
F. Corberi, G. Gonnella, and A. Lamura,
Two-scale competition in phase separation with shear,
Physical Review Letters {\bf 83} (1999) 4057 - 4060.

\bibitem{badal}
V. E. Badalassi, H. D. Ceniceros and S. Banerjee, Gravitational effects on 
structure
development in quenched complex fluids, Annals of the New York Academy of
Sciences {\bf 1027} (2004) 371 - 382.


\end{thebibliography}
\end{document}